\documentclass[aps,prd,twocolumn,superscriptaddress,nofootinbib]{revtex4-2}
\usepackage{amsmath,amssymb,graphicx,float,hyperref}
\usepackage[utf8]{inputenc}
\usepackage{multirow}
\begin{document}

\title{Virtual Majorana Neutrinos and the Minimum Neutrino Mass Scale in Neutrinoless Double-Beta Decay}

\author{Dongming Mei}
\author{Kunming Dong}
\author{Austin Warren}
\author{Sanjay Bhattarai}
\affiliation{University of South Dakota, 414 East Clark Street, Vermillion, South Dakota, USA}
\date{\today}

\begin{abstract}
Virtual Majorana neutrinos are indispensable for neutrinoless double-beta (\(0\nu\beta\beta\)) decay. In this study, we demonstrate that the overlap of the virtual Majorana neutrino wavefunction—predominantly composed of a right-handed antineutrino component with a strongly suppressed left-handed component (with amplitude proportional to the effective Majorana neutrino mass, \(|m_{\beta\beta}|\))—is crucial for triggering this decay process. This effective mass, derived from the minimum neutrino mass, offers valuable insights into the absolute neutrino mass scale. Using best-fit parameters from neutrino oscillation experiments, the minimum neutrino mass is determined from the sum of the three neutrino mass eigenstates,
$\Sigma = m_1 + m_2 + m_3,$
which is represented by two narrow bands centered at approximately \(0.06\,\mathrm{eV/c^2}\) for the normal hierarchy (NH) and \(0.102\,\mathrm{eV/c^2}\) for the inverted hierarchy (IH). Under these constraints, the minimum neutrino mass is found to be \(0.001186\,\mathrm{eV/c^2}\) for NH and \(0.002646\,\mathrm{eV/c^2}\) for IH, thereby establishing a potential absolute neutrino mass scale for both scenarios. From these values, we calculate \(|m_{\beta\beta}|\), which plays a central role in \(0\nu\beta\beta\) decay. By combining \(|m_{\beta\beta}|\) with decay phase-space factors, nuclear matrix elements, and the absorption probability of the virtual Majorana neutrino, we estimate the \(0\nu\beta\beta\) half-life for key isotopes—namely, \(^{76}\mathrm{Ge}\), \(^{130}\mathrm{Te}\), and \(^{136}\mathrm{Xe}\)—using two independent methods. The results are in good agreement, and we also discuss the uncertainties in the nuclear matrix elements that may affect these calculations.

\end{abstract}

\maketitle

\section{Introduction}

Neutrinoless double-beta (\(0\nu\beta\beta\)) decay is a highly anticipated process in particle physics that, if observed, would provide groundbreaking insights into the nature of neutrinos—specifically, confirming that they are Majorana particles and that lepton number is violated by two units \cite{Furry1939}. In contrast to standard double-beta decay (\(2\nu\beta\beta\)), which emits two electrons and two antineutrinos,  the \(0\nu\beta\beta\) decay occurs without antineutrino emission.

The observation of \(0\nu\beta\beta\) decay implies that neutrinos are Majorana particles \cite{Majorana1937,Schechter1982}, directly challenging the Standard Model assumption that neutrinos are Dirac particles with distinct neutrino and antineutrino identities. For \(0\nu\beta\beta\) decay to occur, an antineutrino emitted by one neutron must be absorbed by another as a neutrino. This process requires both an identity change and a helicity flip —outcomes that are only possible if the neutrino is its own antiparticle (i.e., a Majorana particle) and if neutrinos possess a nonzero mass \cite{Bilenky2015,Kayser1982}. 
If neutrinos are Majorana particles, the identity transition occurs naturally, and the probability of the required helicity flip is proportional to the effective Majorana neutrino mass, \(|m_{\beta\beta}|\)~\cite{Kayserpdg}. Thus, provided neutrinos have nonzero mass, this condition is inherently satisfied.

Helicity—the projection of a particle's spin along its direction of motion—is crucial in weak interactions, which couple only to left-handed neutrinos and right-handed antineutrinos. Consequently, for \(0\nu\beta\beta\) decay to proceed, the predominantly right-handed antineutrino must flip to become left-handed. This helicity transition is highly suppressed by the small neutrino mass, rendering the decay exceedingly rare. Precise knowledge of neutrino masses, as established by numerous neutrino oscillation experiments, is therefore essential to understand the scale of this suppression.

Neutrino oscillation experiments have unequivocally demonstrated that neutrinos possess mass and mix among flavor states, a discovery that has significantly challenged the Standard Model. Landmark experiments—such as SNO \cite{SNO2002}, KamLAND \cite{KamLAND2003}, Super-Kamiokande \cite{SuperK2018}, Daya Bay \cite{DayaBay2012}, RENO \cite{RENO2012}, Double Chooz \cite{DoubleChooz2012}, NOvA \cite{NOvA2021}, T2K \cite{T2K2022}, Borexino \cite{Borexino2018}, MINOS \cite{MINOS2013}, K2K \cite{K2K2006}, and CHOOZ \cite{CHOOZ1999}—have provided a robust framework for understanding neutrino oscillations. Despite these successes, critical questions remain: What is the absolute neutrino mass scale? Are neutrinos Dirac or Majorana particles? These issues continue to drive both experimental and theoretical research \cite{PDG2023}.

By correlating the measured mass-squared differences with the sum of the neutrino masses (\(\Sigma = m_1 + m_2 + m_3\)) as a function of the lightest neutrino mass, one can determine the minimum neutrino mass for both the normal hierarchy (NH) and inverted hierarchy (IH) scenarios. Recent analyses reveal two distinct narrow bands centered at \(\Sigma \sim 0.06~\mathrm{eV/c^2}\) for NH and \(\Sigma \sim 0.102~\mathrm{eV/c^2}\) for IH \cite{Mei2024,huang,raul,cao,bura,peter}, consistent with cosmological constraints from Planck observations and other studies \cite{Planck2018,Tristan2022}. These results not only highlight the role of oscillation experiments in constraining neutrino masses but also provide a unified picture of neutrino mass hierarchies.

Furthermore, recent studies have leveraged oscillation parameters to derive the effective Majorana neutrino mass, \(|m_{\beta\beta}|\), which is a critical parameter for \(0\nu\beta\beta\) decay \cite{Furry1939,Avignone2008}. The intrinsic connection between the minimum neutrino mass and \(|m_{\beta\beta}|\) underscores the importance of precise measurements in unraveling neutrino properties. By combining \(|m_{\beta\beta}|\) with decay phase-space factors and nuclear matrix elements, one can estimate the \(0\nu\beta\beta\) half-lives for isotopes such as \(^{76}\mathrm{Ge}\), \(^{130}\mathrm{Te}\), and \(^{136}\mathrm{Xe}\).

In this paper, we investigate the mechanism through which a right-handed antineutrino undergoes a helicity flip, effectively transitioning into a left-handed neutrino during propagation as a virtual Majorana neutrino. Unlike Dirac neutrinos, Majorana neutrinos inherently possess both right-handed and left-handed helicity components. In the \(0\nu\beta\beta\) decay process, a right-handed antineutrino emitted at one vertex propagates to another vertex, where it can be absorbed only by the left-handed current of the Standard Model. The amplitude of this left-handed absorption component is suppressed by a factor proportional to \(\frac{|m_{\beta\beta}|c^2}{E}\), where \(E\) denotes an energy term associated with the virtual Majorana neutrino. Consequently, the helicity-flip probability at the absorption vertex is directly proportional to \(\left(\frac{m_{\beta\beta}c^2}{E}\right)^2\). Therefore, despite the significant suppression associated with a small effective Majorana neutrino mass, the amplitude for the \(0\nu\beta\beta\) decay remains viable and is fundamentally related to \(|m_{\beta\beta}|\).

To calculate \(|m_{\beta\beta}|\), it is essential first to determine the minimum neutrino mass, which can be derived from the sum of the three neutrino mass eigenstates using the measured squared mass differences from neutrino oscillation experiments. Accordingly, in Section~\ref{sec:II} we review the latest neutrino oscillation parameters and derive the minimum neutrino masses for both the NH and IH scenarios. From these values, we compute the corresponding effective Majorana mass, \(|m_{\beta\beta}|\), and explore its implications for \(0\nu\beta\beta\) decay in Section~\ref{sec:III}. By combining \(|m_{\beta\beta}|\) with decay phase-space factors and nuclear matrix elements, we then estimate the \(0\nu\beta\beta\) half-lives for key isotopes, such as \(^{76}\mathrm{Ge}\), \(^{130}\mathrm{Te}\), and \(^{136}\mathrm{Xe}\), and discuss the uncertainties in the nuclear matrix elements that may affect these calculations in Section~\ref{sec:IV}. In Section~\ref{sec:V}, we also examine the strongly suppressed absorption probability of the left-handed neutrino for a given \(|m_{\beta\beta}|\) and its implications. Overall, our analysis, summarized in Section~\ref{sec:VI}, underscores that \(0\nu\beta\beta\) decay searches provide a direct test of the Majorana nature of neutrinos and lepton-number violation, offering new insights into neutrino masses and physics beyond the Standard Model.

\section{The Role of Virtual Majorana Neutrinos}
\label{sec:I}
The \(0\nu\beta\beta\) decay process offers a unique window into the Majorana nature of neutrinos. In this rare event, two neutrons within a nucleus transform into two protons, accompanied by the emission of two electrons. The process is mediated by a virtual Majorana neutrino, whose distinct property as its own antiparticle enables the necessary transformation. A crucial requirement for \(0\nu\beta\beta\) decay is that the antineutrino emitted at one vertex must be absorbed as a neutrino at the other. This is only possible if neutrinos are virtual Majorana particles, allowing the helicity flip required during propagation from one vertex to the other.

\begin{figure}[H]
    \centering
    \includegraphics[width=0.9\linewidth]{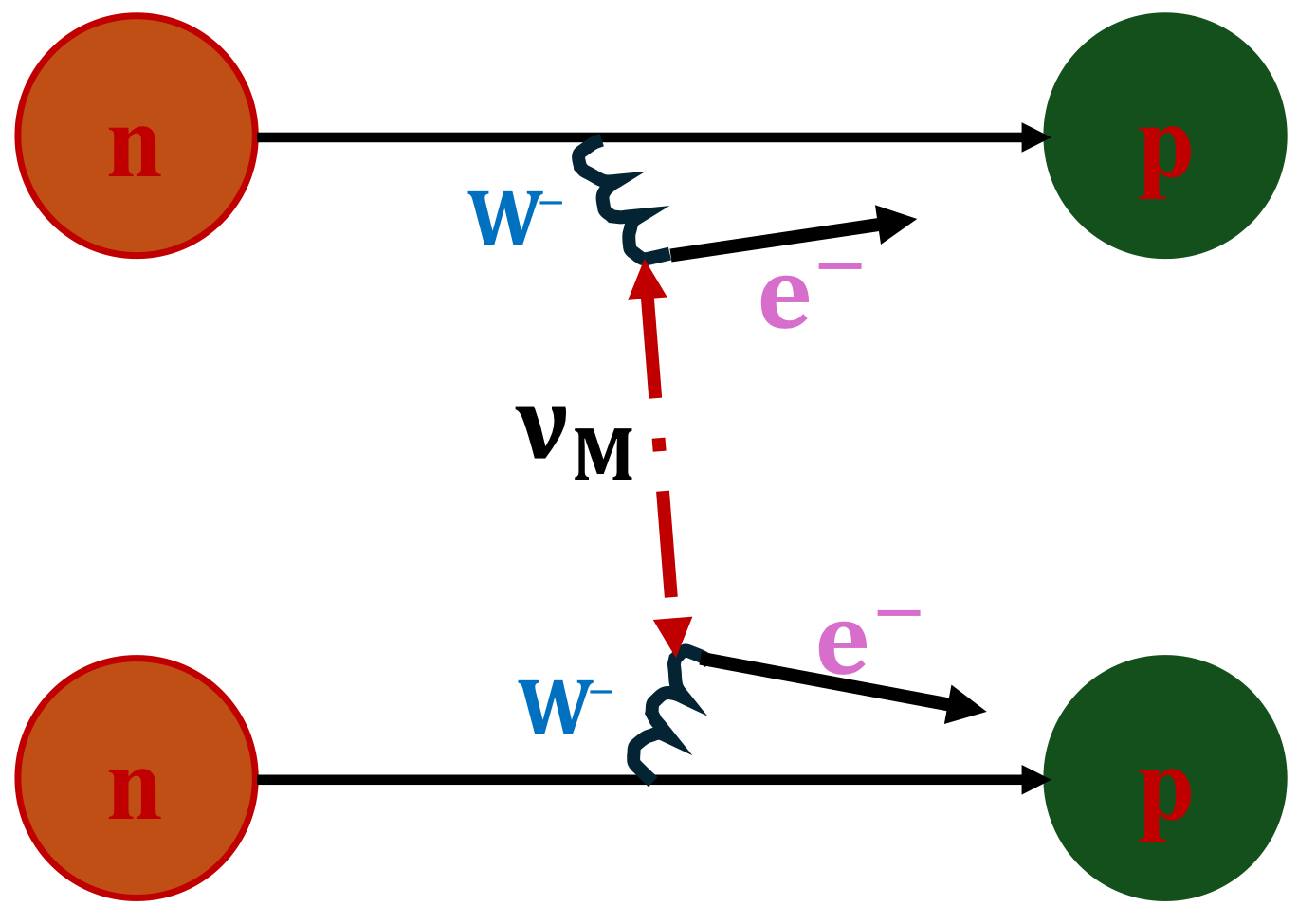}
    \caption{Feynman diagram of \(0\nu\beta\beta\) decay process, mediated by a virtual Majorana neutrino. The process involves two neutrons converting into two protons with the emission of two electrons, where the virtual Majorana neutrino acts as the mediator.}
    \label{fig:majorana_decay}
\end{figure}

Figure~\ref{fig:majorana_decay} illustrates the \(0\nu\beta\beta\) decay process. At one vertex, a neutron transforms into a proton, emitting a \(W^{-}\) boson, which immediately decays into an electron and a right-handed antineutrino (a virtual Majorana neutrino, \(\nu_M\)). This \(\nu_M\) is then absorbed as a left-handed electron neutrino at a second vertex, facilitating the decay of another neutron. For this process to occur, the right-handed antineutrino must convert into a left-handed neutrino—a transition that is only possible if neutrinos and antineutrinos are identical (i.e., if they are Majorana particles) and if the helicity can flip during propagation. One natural explanation for this phenomenon is the overlapping of the neutrino wavefunctions. If Majorana neutrinos possess mass, their helicity can flip during wavefunction propagation, thereby enabling the \(0\nu\beta\beta\) decay process through the overlapping wavefunctions of a virtual Majorana neutrino.

The overlapping wavefunction for a Majorana neutrino, including both left-handed and right-handed components, can be expressed as:
\[
\Psi_M (x) = \Psi_L (x) + \Psi_R (x),
\]
where:
\( \Psi_M (x) \) is the total Majorana neutrino wavefunction, \( \Psi_L (x) \) is the left-handed component, \( \Psi_R (x) \) is the right-handed component.

\begin{figure}[H]
    \centering
    \includegraphics[width=1.0\linewidth]{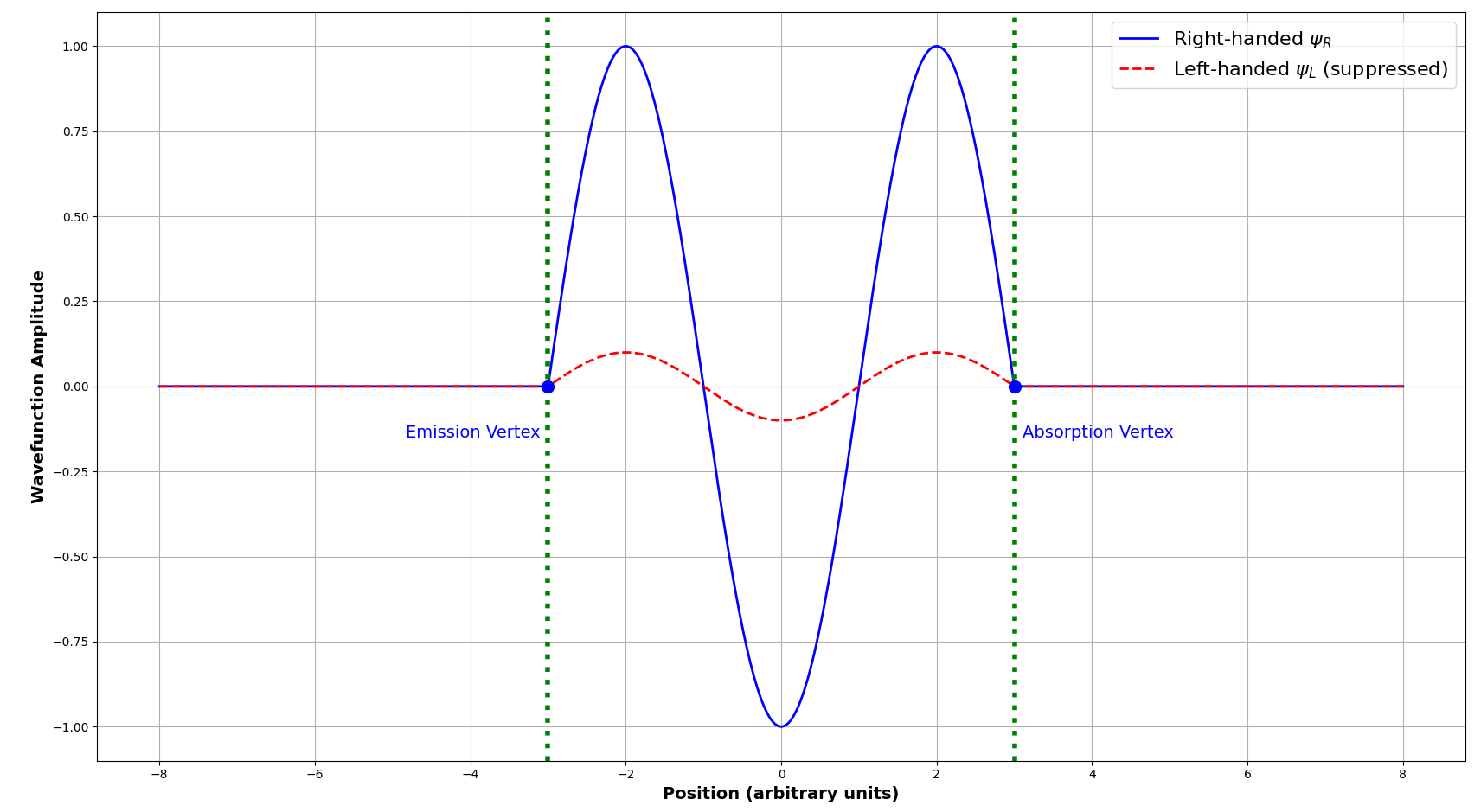}
    \caption{Illustration of wavefunction overlap at two vertices in quantum-entangled states, depicting the mechanism for helicity flip in \(0\nu\beta\beta\) decay. (Note: The wavefunction components are not drawn to scale.)
}
    \label{fig:wavefunction}
\end{figure}

In \(0\nu\beta\beta\) decay, the virtual Majorana neutrino is predominantly right-handed, with a small left-handed component whose amplitude is proportional to \( \frac{|m_{\beta\beta}|c^2}{E} \)~\cite{Kayserpdg}. Therefore, the overlapping wavefunction can be written as:
\[
\Psi_M (x) = \left( \frac{|m_{\beta\beta}|c^2}{E} \right) \Psi_L (x) + \Psi_R (x).
\]
This formulation naturally accounts for the suppression of the left-handed component, which is crucial for the helicity flip mechanism in \(0\nu\beta\beta\) decay. In essence, no explicit physical helicity flip is required; the overlapping wavefunctions allow the predominantly right-handed antineutrino emitted at one vertex to be absorbed at another vertex through its small left-handed component. Figure~\ref{fig:wavefunction} is a schematic plot illustrating the overlap of wavefunctions—modeled as sine functions—at two vertices that are in quantum-entangled states.

The overlapping of virtual Majorana neutrino wavefunctions can be understood by applying the Heisenberg Uncertainty Principle:
\[
\Delta E\, \Delta t \sim \hbar,
\]
where \(\Delta E\) is the energy uncertainty of the virtual neutrino, \(\Delta t\) is the lifetime of the virtual state, and \(\hbar\) is the reduced Planck constant. In \(0\nu\beta\beta\) decay, the virtual neutrino is exchanged between two nucleons, which are separated by a distance comparable to the nuclear diameter. The nuclear diameter is given by \( d = 2R \), where \( R \approx r_{0} A^{1/3} \) with \( r_{0} = 1.2 \times 10^{-15} \,\text{m} \) as the nuclear radius and \( A \) as the atomic mass number. 

For specific isotopes, the estimated separation distances are approximately:
\( d(^{76}\text{Ge}) \approx 1.01 \times 10^{-14} \,\text{m} \),
\( d(^{130}\text{Te}) \approx 1.21 \times 10^{-14} \,\text{m} \), and
\( d(^{136}\text{Xe}) \approx 1.23 \times 10^{-14} \,\text{m} \).

  Using the relation  
\[
\Delta E \sim \frac{\hbar c}{d},
\]
and substituting \( c = 3.0 \times 10^8\,\text{m/s} \) and \( \hbar = 6.58 \times 10^{-16}\,\text{eV} \cdot \text{s} \), we obtain:

\[
\Delta E \approx 19.5\,\text{MeV}, \quad 16.3\,\text{MeV}, \quad \text{and} \quad 16.1\,\text{MeV}
\]

for \(^{76}\text{Ge}\), \(^{130}\text{Te}\), and \(^{136}\text{Xe}\), respectively.

Thus, the lifetime of the virtual neutrino is estimated as  
\[
\Delta t \sim \frac{\hbar}{\Delta E},
\]
which gives  
\[
\Delta t \approx 3.4 \times 10^{-23}\,\text{s}, \quad 4.0 \times 10^{-23}\,\text{s}, \quad \text{and} \quad 4.1 \times 10^{-23}\,\text{s}
\]
for \(^{76}\text{Ge}\), \(^{130}\text{Te}\), and \(^{136}\text{Xe}\), respectively.

This extremely short interval is consistent with the time required for a neutrino to propagate between two nucleons within a nucleus, thereby enabling the decay process without violating energy conservation.

An atomic nucleus is a densely packed system of nucleon fermions (protons and neutrons) bound together by the strong nuclear force. The motion of these nucleons is governed by Fermi-Dirac statistics, and the nuclear Fermi momentum determines the energy levels of these particles near the Fermi surface. Consequently, momentum transfer within the nucleus is primarily dictated by the nuclear Fermi momentum (\(p_F\)), which represents the typical momentum of a nucleon. This momentum can be estimated using the empirical formula~\cite{fermimomentum}.

\begin{equation}
p_F = ( 0.27 - 1.13/A + 9.73/A^2 - 39.53/A^3),
\end{equation}

where $ p_F $ is in GeV, and \( A \) is the atomic mass number.

The neutron Fermi momentum is then calculated using~\cite{fermimomentum}:

\begin{equation}
    p_F^{n} = p_F \left(\frac{2(A-Z)}{A}\right)^{1/3},
\end{equation}

where $ Z $ is the atomic number.  

Since the exchanged neutrino in 0$\nu\beta\beta$ decay is off-shell, its energy is expected to be comparable to the nuclear momentum scale. This can be estimated as:

\begin{equation}
E_{\nu_M} \approx \sqrt{(p_F^{n})^2 c^2 + m_{\beta\beta}^2 c^4}.
\end{equation}

Due to the many orders of magnitude difference between \( p_{F}^{n} \) and \( |m_{\beta\beta}| \), the contribution from \( |m_{\beta\beta}| \) is negligible. Thus, we approximate the energy as:

\begin{equation}
E_{\nu_M} \approx p_{F}^{n}c.
\end{equation}

Substituting the values for specific isotopes, we obtain:

\[
\begin{array}{rcl}
E_{\nu_M}(^{76}\text{Ge}) &\approx& 269.6\,\text{MeV}, \\
E_{\nu_M}(^{130}\text{Te}) &\approx& 278.2\,\text{MeV}, \\
E_{\nu_M}(^{136}\text{Xe}) &\approx& 278.9\,\text{MeV}.
\end{array}
\]

Since the energy of the virtual Majorana neutrino involved in $0\nu\beta\beta$ decay is approximately $270\,\text{MeV}$ and its lifetime is on the order of $ 10^{-23}\,\text{s}$, the wavefunction overlap required for helicity flip collapses within an extremely short time scale. This suggests that the virtual Majorana neutrino could propagate at a speed far exceeding that of light in vacuum. A natural explanation for this apparent superluminal behavior is that the virtual particle exists in a quantum-entangled state, unconstrained by the conventional energy-momentum relation. In this state, the overlap between its right- and left-handed components naturally facilitates the helicity flip essential for the $0\nu\beta\beta$ decay process.

This quantum behavior is intimately connected to the fundamental nature of Majorana neutrinos. Unlike Dirac neutrinos, which have distinct particle and antiparticle states, Majorana neutrinos possess mass and inherently incorporate both left-handed and right-handed components within a single wavefunction. In \(0\nu\beta\beta\) decay, a right-handed antineutrino emitted at one vertex must be absorbed at another vertex via its small left-handed component through the Standard Model's left-handed current. Analogous to Dirac neutrino oscillations—where the oscillation probability depends on the mass-squared difference between two mass eigenstates, the travel distance, and the neutrino energy—we can similarly construct the probability for the absorption of a virtual Majorana neutrino between two neutrons within the nucleus. This absorption probability is approximately proportional to
\[
\frac{(m_{\beta\beta}c^2)^2\,d}{\hbar c\,E_{\nu_M}},
\]
where \(\Delta E = \frac{\hbar c}{d}\) as described above. Assuming that the effective absorption region of the virtual Majorana neutrino can be approximated as a sphere, this probability can be expressed as:

\begin{equation}
\label{eq:8}
P = 4\pi\left(\frac{(m_{\beta\beta}c^2)^2}{\Delta E\cdot E_{\nu_M}}\right).
\end{equation}
Thus, for a virtual Majorana neutrino with a small effective Majorana mass (e.g., \(|m_{\beta\beta}| \sim 0.01\,\text{eV/c$^2$}\)), the absorption process is highly suppressed at typical energy scales.

At first glance, this suppression might seem to imply that the virtual neutrino could travel superluminally to account for the observed decay rate. However, within the framework of quantum field theory, virtual particles are not constrained by the usual energy-momentum relation, allowing them to exhibit such apparent behavior without violating causality. In particular, the quantum entanglement of the neutrino wavefunction enables interference effects that facilitate the necessary absorption of the left-handed neutrino, thereby allowing the $0\nu\beta\beta$ decay process to occur.

The ability of a right-handed antineutrino to be absorbed as a left-handed neutrino is a direct consequence of the Majorana nature of neutrinos, linking neutrino mass, lepton-number violation, and the fundamental symmetries of nature. This process is in stark contrast to Dirac neutrinos, where lepton number conservation prohibits such a transition \cite{Majorana1937,Mohapatra2007}. The overlap of the right- and left-handed components at the decay vertices provides an elegant mechanism for $0\nu\beta\beta$ decay \cite{Schechter1982}, reinforcing the significance of measuring the effective Majorana mass, $|m_{\beta\beta}|$. As discussed in Sections~\ref{sec:III} and \ref{sec:IV}, precise measurements of $0\nu\beta\beta$ decay play a crucial role in probing the absolute neutrino mass scale and exploring physics beyond the Standard Model. These studies offer deep insights into the origin of neutrino masses, the matter-antimatter asymmetry of the universe, and the fundamental symmetries governing particle interactions.

The effective Majorana mass, \(|m_{\beta\beta}|\), which quantifies the amplitude of the left-handed neutrino wavefunction as part of the virtual Majorana neutrino wavefunction, depends on the mixing matrix elements (\(U_{ei}\)) and the masses of the neutrino eigenstates (\(m_1, m_2, m_3\)):
\begin{equation}
\label{eq:eq1}
|m_{\beta\beta}| = \left| \sum_{i=1}^3 U_{ei}^2 m_i \right|.
\end{equation} 

The nonzero masses of these eigenstates are crucial for maintaining the coherence and overlap of wavefunctions required for \(0\nu\beta\beta\) decay. Ultimately, the observation of \(0\nu\beta\beta\) decay would not only confirm the Majorana nature of neutrinos and the violation of lepton number but also provide vital insights into the absolute neutrino mass scale.

\section{Calculation of the Minimum Neutrino Mass}
\label{sec:II}
The determination of the minimum neutrino mass, \( m_L \), relies on the measured best-fit parameters from neutrino oscillation experiments, which provide the squared mass differences and mixing angles. The most recent measurements reported in the \textit{Particle Data Group 2022} (PDG 2022) \cite{PDG2023} are as follows:

\begin{itemize}
    \item \textbf{Normal Hierarchy (NH):}
    \[
    \begin{array}{rcl}
        \Delta m^2_{21} &= (7.41^{+0.21}_{-0.20}) \times 10^{-5}\,\mathrm{eV}^2/c^4, \\
        \Delta m^2_{32} &= (2.437^{+0.028}_{-0.027}) \times 10^{-3}\,\mathrm{eV}^2/c^4, \\
        \theta_{13}    &= (8.54_{-0.12}^{+0.11})^\circ, \\
        \sin^2\theta_{13} &= (2.203_{-0.059}^{+0.056}) \times 10^{-2}, \\
        \theta_{23}    &= (49.1_{-1.3}^{+1.0})^\circ, \\
        \sin^2\theta_{23} &= (5.71_{-0.23}^{+0.18}) \times 10^{-1}.
    \end{array}
    \]
    \item \textbf{Inverted Hierarchy (IH):}
    \[
    \begin{array}{rcl}
        \Delta m^2_{21} &= (7.41^{+0.21}_{-0.20}) \times 10^{-5}\,\mathrm{eV}^2/c^4, \\
        \Delta m^2_{32} &= (-2.498^{+0.032}_{-0.025}) \times 10^{-3}\,\mathrm{eV}^2/c^4, \\
        \theta_{13}    &= (8.57_{-0.11}^{+0.12})^\circ, \\
        \sin^2\theta_{13} &= (2.219_{-0.057}^{+0.060}) \times 10^{-2}, \\
        \theta_{23}    &= (49.5_{-1.2}^{+0.9})^\circ, \\
        \sin^2\theta_{23} &= (5.78_{-0.21}^{+0.16}) \times 10^{-1}.
    \end{array}
    \]
    \item \textbf{Mixing Angles (common for NH and IH):}
    \[
    \begin{array}{rcl}
        \theta_{12}    &= (33.41_{-0.72}^{+0.75})^\circ, \\
        \sin^2\theta_{12} &= (3.03_{-0.11}^{+0.12}) \times 10^{-1}.
    \end{array}
    \]
\end{itemize}

\begin{figure}[H]
    \centering
    \includegraphics[width=1.0\linewidth]{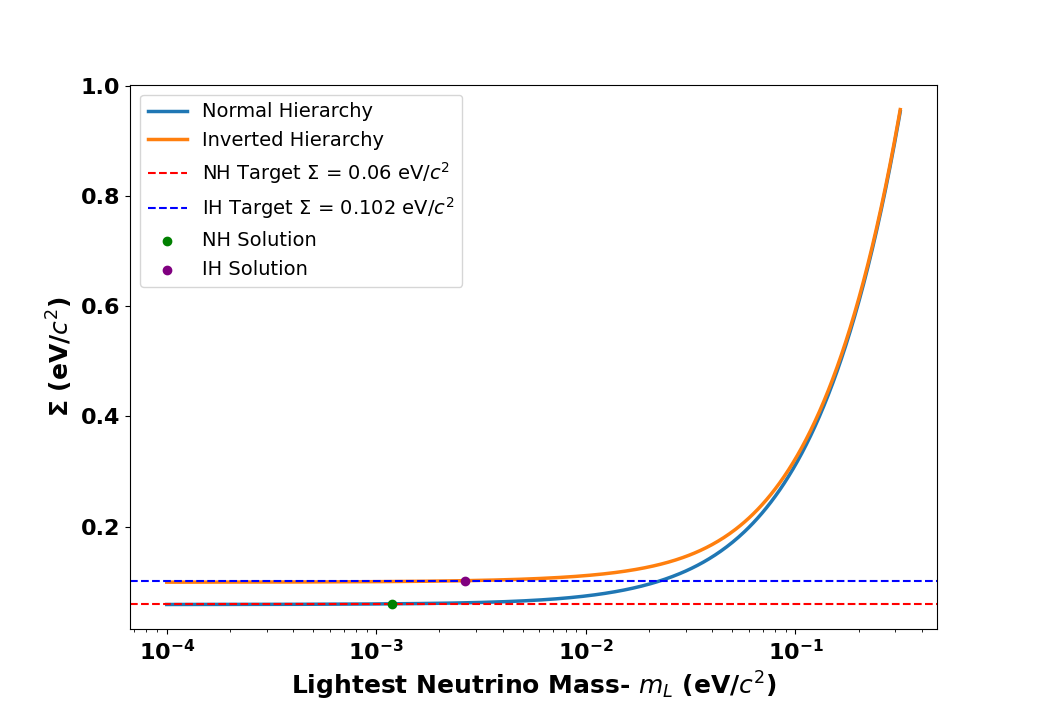}
    \caption{The sum of neutrino masses, \(\Sigma = m_1 + m_2 + m_3\), as a function of the minimum neutrino mass, \(m_L\), for both NH and IH scenarios. The bands reflect the constants when the minimum neutrino mass is less than 3$\times10^{-3} eV/c^2$, with centers at \(0.06~\mathrm{eV/c^2}\) for NH and \(0.102~\mathrm{eV/c^2}\) for IH. The intersections of these bands with the curves determine the minimum neutrino mass for each hierarchy.}
    \label{fig:Sigma_vs_mL}
\end{figure}

The sum of neutrino masses, \(\Sigma = m_1 + m_2 + m_3\), can be expressed as a function of the lightest neutrino mass, \(m_L\), using the measured mass-squared differences from neutrino oscillation experiments. Figure~\ref{fig:Sigma_vs_mL} illustrates this relationship, showing two distinct narrow bands on \(\Sigma\): \(0.06~\mathrm{eV/c^2}\) for NH and \(0.102~\mathrm{eV/c^2}\) for IH. For NH, \(\Sigma\) corresponds to a lightest mass dictated by \(m_1\), whereas for IH, \(\Sigma\) is dictated by \(m_3\). These constraints enable precise determination of the minimum neutrino mass by intersecting the \(\Sigma\) curve with the two narrow bands. The functional dependence of \(\Sigma\) on \(m_L\) is given by:

\begin{itemize}
    \item For NH (\(m_1 < m_2 < m_3\)):
    \begin{eqnarray}
        \Sigma_\mathrm{NH} &=& m_1 + \sqrt{m_1^2 + \Delta m^2_{21}} \nonumber \\
        && + \sqrt{m_1^2 + \Delta m^2_{32}}.
    \end{eqnarray}
    \item For IH (\(m_3 < m_1 < m_2\)):
    \begin{eqnarray}
        \Sigma_\mathrm{IH} &=& m_3 + \sqrt{m_3^2 - \Delta m^2_{32}} \nonumber \\
        && + \sqrt{m_3^2 - \Delta m^2_{32} + \Delta m^2_{21}}.
    \end{eqnarray}
\end{itemize}

As shown in Figure~\ref{fig:Sigma_vs_mL}, two narrow bands: \(0.06~\mathrm{eV/c^2}\) for NH and \(0.102~\mathrm{eV/c^2}\) for IH, are nearly constants when the minimum neutrino mass is less than 3$\times10^{-3} eV/c^2$. The intersections of these bands with the \(\Sigma\) curves determine \(m_L\). The iterative numerical process, implemented through computational code, ensures that the computed 
\( \Sigma \)
matches the chosen band within acceptable precision, yielding the minimum neutrino mass and the full set of eigenstate masses for both hierarchies. This solution offers a potential absolute neutrino mass scale when the sum of the three mass eigenstates is accurately determined.

Using the central values of \(\Sigma\), the corresponding minimum neutrino masses are presented in Table~\ref{tab:minimum}:

\begin{table}[h]
    \centering
    \renewcommand{\arraystretch}{1.3}
    \begin{tabular}{|c|c|c|c|}
        \hline
        \textbf{Hierarchy} & \textbf{Mass State} & \textbf{Expression}& \textbf{Value (eV/c$^2$)} \\
        \hline
        \multirow{3}{*}{NH} 
        & $m_1$ & - & 0.001186 \\
        & $m_2$ & $\sqrt{m_1^2 + \Delta m^2_{21}}$ & 0.008689 \\
        & $m_3$ & $ \Sigma - m_1 - m_2 $ & 0.050125 \\
        \hline
        \multirow{3}{*}{IH} 
        & $m_3$ & - & 0.002646 \\
        & $m_1$ & $\Sigma - m_3 - m_2$ & 0.049304 \\
        & $m_2$ & $\sqrt{m_3^2 + \Delta m^2_{21}}$ & 0.050050 \\
        \hline
    \end{tabular}
    \caption{Neutrino mass eigenstates for the Normal and Inverted Hierarchies. The values are calculated using best-fit neutrino oscillation parameters.}
    \label{tab:minimum}
\end{table}

Figure~\ref{fig:Sigma_vs_mL} illustrates the relationship between \(\Sigma\) and \(m_L\), providing insight into the minimum neutrino mass, a critical input for understanding the effective Majorana mass and 0$\nu\beta\beta$ decay.

It is worth mentioning that in the NH scenario, the calculated neutrino masses indicate that \( m_3 \) is significantly larger than \( m_1 \) and \( m_2 \), resulting in a well-separated mass eigenstate. Conversely, in the IH scenario, \( m_1 \) and \( m_2 \) are nearly degenerate, with \( m_3 \) being much smaller. The lightest masses—\( m_1 \) in NH and \( m_3 \) in IH—fall within the ranges reported in previous studies by Mei et al.~\cite{Mei2024}, corroborating earlier findings.

The nonzero masses of the eigenstates (\(m_1, m_2, m_3\)) define a possible scenario for the absolute neutrino mass scale, ensuring the coherence and constructive interference of their wavefunctions, which in turn generates a finite \(|m_{\beta\beta}|\). This effective mass directly influences the amplitude of the \(0\nu\beta\beta\) process, establishing a connection between the quantum mechanical properties of neutrinos and an experimentally measurable phenomenon. Consequently, the interplay between neutrino masses and wavefunction overlaps provides a natural explanation for the occurrence of \(0\nu\beta\beta\) decay, making it a crucial probe for the Majorana nature of neutrinos and lepton number violation.

\section{Effective Majorana Mass}
\label{sec:III}
The effective Majorana neutrino mass, \( |m_{\beta\beta}| \), is a critical parameter in \(0\nu\beta\beta\) decay searches, as it directly links the neutrino mass spectrum to the intrinsic nature of neutrinos via Eq.~\ref{eq:eq1}. Determining the minimum neutrino mass, \( m_L \), has a significant impact on \( |m_{\beta\beta}| \) by establishing a nonzero lower bound—an essential condition for observing \(0\nu\beta\beta\) decay. Consequently, a nonzero \( |m_{\beta\beta}| \) implies that, if neutrinos are Majorana particles, the decay may be detectable in planned ton-scale or even 100-ton scale next-generation experiments~\cite{Rodejohann2011, Mei2024}.

To compute \( |m_{\beta\beta}| \), we follow the formalism outlined in Mei et al.~\cite{Mei2024} using the method developed by Zhi-Zhong Xing and Ye-Ling Zhou~\cite{xing}.  
For NH, where \( m_1 < m_2 < m_3 \), the maximum and minimum values of \( |m_{\beta\beta}| \) are given by:

\begin{align}
    |m_{\beta\beta}|_{\text{max}} &= \sqrt{\Delta m_{21}^2 + m_1^2} \sin^2\theta_{12} \cos^2\theta_{13} \Bigg[ 1 + \notag \\ 
    &\quad \sqrt{1 - \frac{\Delta m_{21}^2}{\Delta m_{21}^2 + m_1^2}} \cot^2\theta_{12} + \notag \\ 
    &\quad \sqrt{1 - \frac{\Delta m_{21}^2}{\Delta m_{21}^2 + m_1^2} + \frac{\Delta m_{21}^2 + \Delta m_{32}^2}{\Delta m_{21}^2 + m_1^2}} \notag \\
    &\quad \times \frac{\tan^2\theta_{13}}{\sin^2\theta_{12}} \Bigg],
\end{align}

\begin{align}
    |m_{\beta\beta}|_{\text{min}}^{(1)} &= \sqrt{\Delta m_{21}^2 + m_1^2} \sin^2\theta_{12} \cos^2\theta_{13} \Bigg[ 1 - \notag \\
    &\quad \sqrt{1 - \frac{\Delta m_{21}^2}{\Delta m_{21}^2 + m_1^2}} \cot^2\theta_{12} - \notag \\ 
    &\quad \sqrt{1 - \frac{\Delta m_{21}^2}{\Delta m_{21}^2 + m_1^2} + \frac{\Delta m_{21}^2 + \Delta m_{32}^2}{\Delta m_{21}^2 + m_1^2}} \notag \\
    &\quad \times \frac{\tan^2\theta_{13}}{\sin^2\theta_{12}} \Bigg],
\end{align}

\begin{align}
    |m_{\beta\beta}|_{\text{min}}^{(2)} &= \sqrt{\Delta m_{21}^2 + m_1^2} \sin^2\theta_{12} \cos^2\theta_{13} \Bigg[ \notag \\
    &\quad \sqrt{1 - \frac{\Delta m_{21}^2}{\Delta m_{21}^2 + m_1^2}} \cot^2\theta_{12} - 1 - \notag \\
    &\quad \sqrt{1 - \frac{\Delta m_{21}^2}{\Delta m_{21}^2 + m_1^2} + \frac{\Delta m_{21}^2 + \Delta m_{32}^2}{\Delta m_{21}^2 + m_1^2}} \notag \\
    &\quad \times \frac{\tan^2\theta_{13}}{\sin^2\theta_{12}} \Bigg].
\end{align}

For IH, where \( m_3 < m_1 < m_2 \), the upper and lower bounds of \( |m_{\beta\beta}| \) are determined by:
\begin{align}
    |m_{\beta\beta}|_{\text{max}} &= \sqrt{m_3^2 - \Delta m_{32}^2} \sin^2\theta_{12} \cos^2\theta_{13} \Bigg[ 1 + \notag \\
    &\quad \sqrt{1 - \frac{\Delta m_{21}^2}{m_3^2 - \Delta m_{32}^2}} \cot^2\theta_{12} + \notag \\
    &\quad \sqrt{1 - \frac{\Delta m_{21}^2}{m_3^2 - \Delta m_{32}^2} + \frac{\Delta m_{21}^2 - \Delta m_{32}^2}{m_3^2 - \Delta m_{32}^2}} \notag \\
    &\quad \times \frac{\tan^2\theta_{13}}{\sin^2\theta_{12}} \Bigg],
\end{align}

\begin{align}
    |m_{\beta\beta}|_{\text{min}} &= \sqrt{m_3^2 - \Delta m_{32}^2} \sin^2\theta_{12} \cos^2\theta_{13} \Bigg[ \notag \\
    &\quad \sqrt{1 - \frac{\Delta m_{21}^2}{m_3^2 - \Delta m_{32}^2}} \cot^2\theta_{12} - 1 - \notag \\
    &\quad \sqrt{1 - \frac{\Delta m_{21}^2}{m_3^2 - \Delta m_{32}^2} + \frac{\Delta m_{21}^2 - \Delta m_{32}^2}{m_3^2 - \Delta m_{32}^2}} \notag \\
    &\quad \times \frac{\tan^2\theta_{13}}{\sin^2\theta_{12}} \Bigg].
\end{align}

Using the best-fit oscillation parameters from PDG 2022~\cite{PDG2023} and the minimum neutrino mass values—$m_1$ for NH and $m_3$ for IH—as determined in Table~\ref{tab:minimum}, we calculate the effective Majorana neutrino mass, $|m_{\beta\beta}|$, for both hierarchies. Table~\ref{tab:mbb_values} displays the results. 

\begin{table}[h]
    \centering
    \renewcommand{\arraystretch}{1.3}
    \begin{tabular}{|c|c|c|}
        \hline
        \textbf{Hierarchy} & \textbf{Parameter} & \textbf{Value (eV/c$^2$)} \\
        \hline
        \multirow{3}{*}{NH} 
        & $|m_{\beta\beta}|_{\text{max}}$ & 0.004488 \\
        & $|m_{\beta\beta}|_{\text{min1}}$ & 0.000662 \\
        & $|m_{\beta\beta}|_{\text{min2}}$ & -0.002871  \\
        \hline
        \multirow{2}{*}{IH} 
        & $|m_{\beta\beta}|_{\text{max}}$ & 0.0500001 \\
        & $|m_{\beta\beta}|_{\text{min}}$ & 0.017204 \\
        \hline
    \end{tabular}
    \caption{Computed values of $|m_{\beta\beta}|$ for Normal and Inverted Hierarchies. The non-physical solution $|m_{\beta\beta}|_{\text{min2}}$ in the NH case is disregarded.}
    \label{tab:mbb_values}
\end{table}

The above results highlight the dependence of \( |m_{\beta\beta}| \) on the hierarchy and the minimum neutrino mass. In the IH scenario, \( |m_{\beta\beta}| \) is significantly larger due to the quasi-degenerate nature of \( m_1 \) and \( m_2 \), making it more favorable for \( 0\nu\beta\beta \) decay detection. However, even in the NH scenario, the determination of a nonzero \( m_L \) ensures a nonzero \( |m_{\beta\beta}| \), providing motivation for experimental searches in this region~\cite{DellOro2016}.

\begin{figure}[H]
    \centering
    \includegraphics[width=1.0\linewidth]{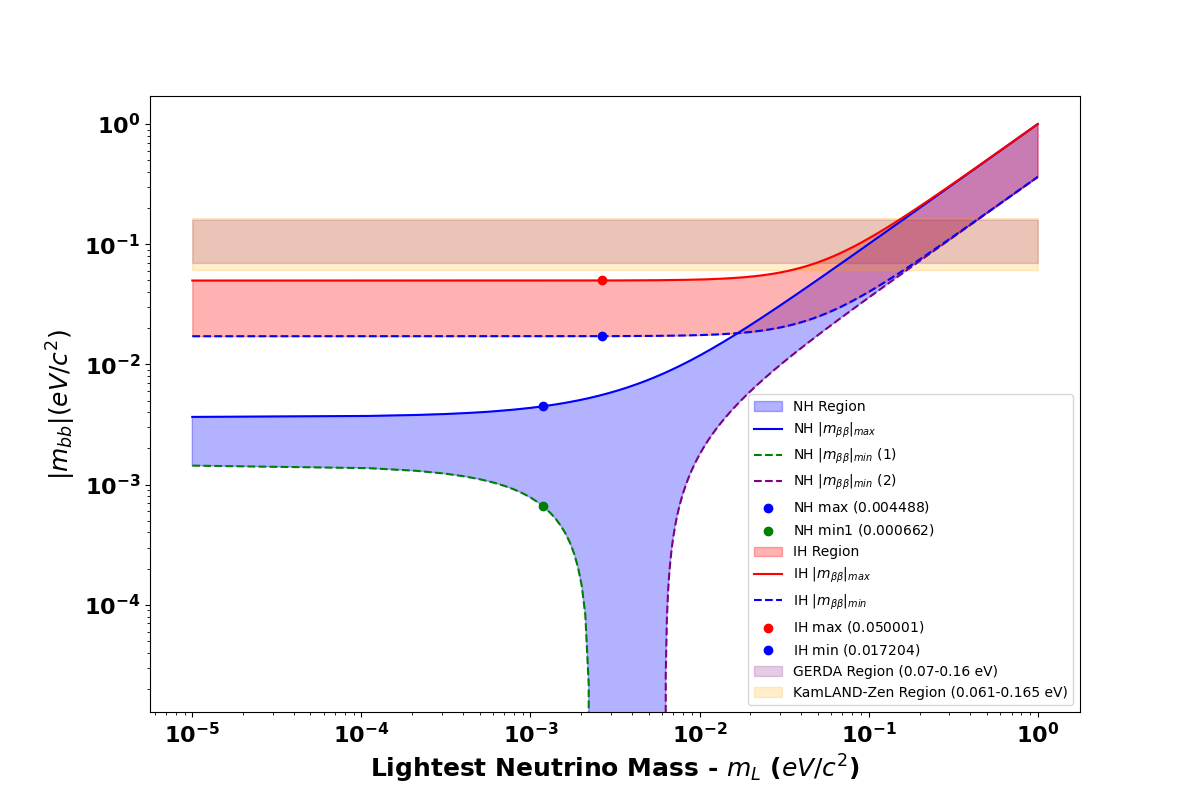}
    \caption{The figure shows \( |m_{\beta\beta}| \) as a function of the minimum neutrino mass, \( m_L \), for both the NH and IH scenarios. It demonstrates that \( |m_{\beta\beta}| \) remains nonzero for the best-fit values of \( m_L \) obtained from oscillation experiments, underscoring the potential detectability of 0$\nu\beta\beta$ decay. The current experimental limits from GERDA~\cite{GERDA2019} and KamLAND-Zen~\cite{KamLANDZen2022} are also displayed.}
    \label{fig:Majorana_mass_vs_mL}
\end{figure}

Figure~\ref{fig:Majorana_mass_vs_mL} illustrates the allowed regions of \( |m_{\beta\beta}| \) as a function of the minimum neutrino mass, \( m_L \). As shown, \( |m_{\beta\beta}| \) remains nonzero for the determined values of \( m_L \) in both hierarchy scenarios, highlighting its potential detectability in \( 0\nu\beta\beta \) decay experiments.

\section{Calculation of $0\nu\beta\beta$ Half-Life for $^{76}$Ge, $^{130}$Te, and $^{136}$Xe}
\label{sec:IV}
The half-life of $0\nu\beta\beta$ is directly related to the effective Majorana neutrino mass $|m_{\beta\beta}|$, the nuclear matrix element ($M_{0\nu}$), and the phase-space factor ($G_{0\nu}$) via the expression~\cite{DellOro2016}:
\begin{equation}
T_{1/2}^{0\nu} = \frac{1}{G_{0\nu} |M_{0\nu}|^2 \left(\frac{|m_{\beta\beta}|}{m_e}\right)^2},
\label{eq:half-life}
\end{equation}
where $m_e$ is the electron mass. 

The $M^{0\nu}$ for $0\nu\beta\beta$ decay is typically expressed as a combination of three components: the Fermi ($M_F$), Gamow-Teller ($M_{GT}$), and Tensor ($M_T$) contributions. The equation is given by~\cite{belley2021}:
\begin{equation}
M^{0\nu} =  \left[ M_{GT} - \left( \frac{g_V}{g_A} \right)^2 M_F + M_T \right],
\end{equation}
where $g_A$ is the axial vector coupling constant, $g_V$ is the vector coupling constant, $M_{GT}$ represents the Gamow-Teller contribution, which involves spin-dependent terms, $M_F$ represents the Fermi contribution, which involves spin-independent terms, $M_T$ represents the Tensor contribution, which accounts for higher-order terms.

The $G^{0\nu}$ for $0\nu\beta\beta$ decay describes the kinematic and geometric aspects of the decay process, involving the emitted electron wavefunctions and the energy released during the decay. The general expression for $G^{0\nu}$ is given by~\cite{Bilenky, Kotila2012}:

\begin{equation}
\begin{array}{c}

G^{0\nu}(Q, Z) = \frac{1}{\ln 2 (2\pi)^5} \frac{(G_F \cos \theta_C)^4}{R^2} \\
\times \int_0^Q E_1 |p_1| E_2 |p_2| F(E_1, Z') F(E_2, Z') \, dT_1,

\end{array}
\end{equation}
where \(G_F\) is the Fermi coupling constant, \(\cos \theta_C\) represents the cosine of the Cabibbo angle, and \(R\) denotes the nuclear radius. The parameter \(Q\) corresponds to the decay \(Q\)-value, which represents the total energy released in the process. The quantities \(E_1\) and \(E_2\) refer to the energies of the emitted electrons, while \(|p_1|\) and \(|p_2|\) denote the magnitudes of their momenta. The term \(F(E, Z')\) is the Fermi function, which accounts for the Coulomb interaction between the outgoing electron and the daughter nucleus. Finally, \(dT_1\) represents the differential kinetic energy of one of the emitted electrons.

The values of  $G_{0\nu}$ and  $M_{0\nu}$ for $^{76}$Ge, $^{130}$Te, and $^{136}$Xe used in this work are summarized in Table~\ref{tab:nuclear_params} \cite{DellOro2016, Dolinski2019, Mei2024}.

\begin{table}[H]
    \centering
    \begin{tabular}{lccc}
        \hline
        Isotope & $G_{0\nu}$ (yr$^{-1}$) & $M_{0\nu}$ & $T_{1/2}^{0\nu}$ (yr) \\
        \hline
        $^{76}$Ge & $2.36 \times 10^{-15}$ & $4.7$ & $1.8 \times 10^{26}$ (GERDA)~\cite{GERDA2019} \\
        $^{130}$Te & $1.45 \times 10^{-14}$ & $3.9$ & $2.2 \times 10^{25}$ (CUORE)~\cite{CUORE2021} \\
        $^{136}$Xe & $1.45 \times 10^{-14}$ & $3.4$ & $2.3 \times 10^{26}$ (KamLAND-Zen)~\cite{KamLANDZen2022} \\
        \hline
    \end{tabular}
    \caption{Nuclear matrix elements ($M_{0\nu}$), phase-space factors ($G_{0\nu}$), and experimental half-life limits for $^{76}$Ge, $^{130}$Te, and $^{136}$Xe.}
    \label{tab:nuclear_params}
\end{table}

\begin{figure}[H]
    \centering
    \includegraphics[width=1.0\linewidth]{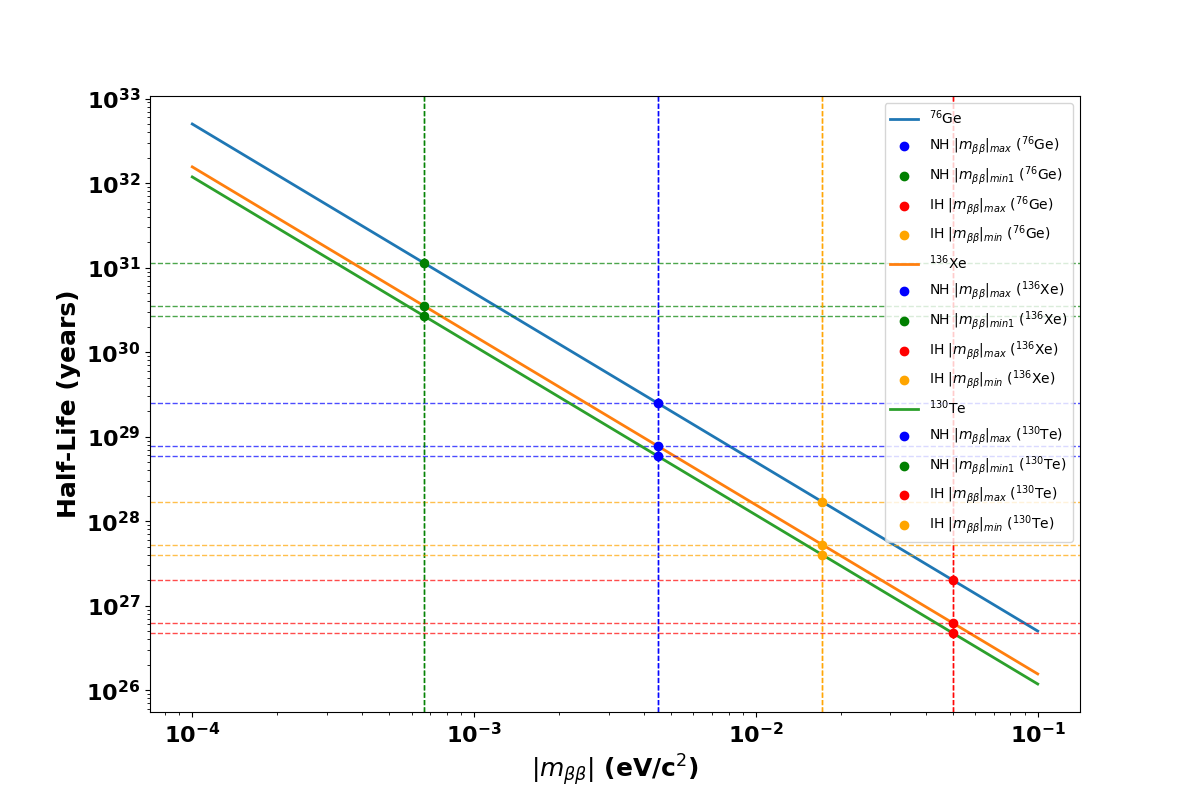}
    \caption{ The figure presents the calculated neutrinoless double-beta decay half-life, \(T_{1/2}^{0\nu}\), as a function of the effective Majorana neutrino mass, $|m_{\beta\beta}|$, for the isotopes \(^{76}\mathrm{Ge}\), \(^{130}\mathrm{Te}\), and \(^{136}\mathrm{Xe}\).
}
    \label{fig:halflife_vs_mL}
\end{figure}

Using the effective Majorana neutrino masses computed in the previous section and applying Equation~\ref{eq:half-life}, we compute the expected \(0\nu\beta\beta\) half-life as a function of the effective Majorana neutrino mass, $|m_{\beta\beta}|$), for both the NH and IH  scenarios. Figure~\ref{fig:halflife_vs_mL} displays the dependence of \(T_{1/2}^{0\nu}\) on \(|m_{\beta\beta}|\) for the three isotopes under consideration.

As seen in Figure~\ref{fig:halflife_vs_mL}, the half-life increases rapidly as $|m_{\beta\beta}|$ decreases. For $^{76}$Ge, $T_{1/2}^{0\nu}$ ranges from $10^{29}$ to $10^{31}$ years for NH and from $10^{27}$ to $10^{28}$ years for IH. Similar trends are observed for $^{130}$Te and $^{136}$Xe, but with differences due to variations in $G_{0\nu}$ and $M_{0\nu}$. These results highlight the importance of precise determinations of $|m_{\beta\beta}|$ and $m_L$ to interpret $0\nu\beta\beta$ decay measurements and distinguish between NH and IH scenarios.

The experimental determination of $|m_{\beta\beta}|$ in $0\nu\beta\beta$ decay is significantly influenced by uncertainties in $G_{0\nu}$ and $M_{0\nu}$ for isotopes such as $^{76}\text{Ge}$, $^{130}\text{Te}$, and $^{136}\text{Xe}$. $G_{0\nu}$, determined by relativistic electron wavefunctions and decay kinematics, relies on precise modeling of Coulomb interactions and atomic effects. Even small variations in $G_{0\nu}$ values can lead to notable discrepancies in decay rate predictions \citep{Kotila2012}.

$M_{0\nu}$, on the other hand, is derived from nuclear structure models, including the Quasiparticle Random-Phase Approximation (QRPA), the shell model, and the interacting boson model. These models introduce methodological uncertainties due to approximations in nuclear correlations, deformation, and interaction effects \citep{Engel2017, Mustonen2013}. Variations in $M_{0\nu}$ values can reach up to 50\% depending on the computational method used.

Combined, these uncertainties directly impact the determination of the effective Majorana neutrino mass ($|m_{\beta\beta}|$), as the half-life is inversely proportional to the product of the squared $G_{0\nu}$ and $M_{0\nu}$. Therefore, accurate and consistent determination of both $G_{0\nu}$ and $M_{0\nu}$ is critical for reliable interpretation of experimental results and for making meaningful comparisons between different isotopes in the quest to observe $0\nu\beta\beta$ decay.

Recent experimental efforts, including GERDA \citep{GERDA2019}, CUORE \citep{CUORE2021}, and KamLAND-Zen \citep{KamLANDZen2022}, provide stringent upper limits on $T_{1/2}^{0\nu}$, which can be used to constrain $G_{0\nu}$ and $M_{0\nu}$ and study the uncertainties of both. Upcoming experiments like LEGEND~\cite{LEGEND2021}, nEXO~\cite{nEXO2022}, and CUPID~\cite{cupid} aim to improve sensitivity by at least an order of magnitude, probing the IH parameter space. These advancements are crucial for testing the Majorana nature of neutrinos and understanding the origin of lepton number violation in the universe.

\section{Discussion and Implications}
\label{sec:V}
 In \(0\nu\beta\beta\) decay, the neutrino involved is virtual rather than real because it must be emitted and absorbed within the nucleus, facilitating the process without appearing in the final state. If the neutrino were real, it would propagate freely and appear in the final state, violating the requirement that no neutrinos are observed in the final state. However, since a Majorana neutrino is its own antiparticle, the neutrino emitted by one neutron can be absorbed by another through overlapping of wavefunctions, effectively closing the loop of the interaction. This requires the neutrino to exist as a virtual intermediate particle, meaning it does not satisfy the usual energy-momentum relation of a free particle but instead acts as a mediator of the weak interaction within the nucleus. The existence of such a virtual Majorana neutrino is crucial for \(0\nu\beta\beta\) decay to occur and provides direct evidence of lepton number violation, which would confirm that neutrinos are Majorana in nature.

In \(0\nu\beta\beta\) decay, the virtual Majorana neutrino is predominantly right-handed, with its left-handed component suppressed by a factor of \(\left(\frac{|m_{\beta\beta}|c^2}{E}\right)\) (see Section~\ref{sec:I}). Consequently, the absorption probability is estimated using Equation~\ref{eq:8}.

For a nucleus with \(N\) neutrons, any pair of neutrons can potentially undergo \(0\nu\beta\beta\) decay mediated by a virtual Majorana neutrino. The number of possible neutron pairs is given by

\begin{equation}
C_X = \frac{N!}{2!(N-2)!},
\end{equation}
where \(X\) denotes the specific nucleus and \(N\) is the total number of neutrons. For instance, the number of possible pairs is 946 for \(^{76}\text{Ge}\), 3003 for \(^{130}\text{Te}\), and 3321 for \(^{136}\text{Xe}\). These values represent the potential number of virtual Majorana neutrino exchanges within the nucleus.

Accordingly, the overall absorption probability of virtual Majorana neutrinos within the nucleus is expressed as:

\begin{equation}
P_X = C_X \cdot 4\pi\,\frac{(m_{\beta\beta} c^2)^2}{\Delta E \cdot E_{\nu_M}}.
\end{equation}

\begin{figure}[h]
    \centering
    \includegraphics[width=1.0\linewidth]{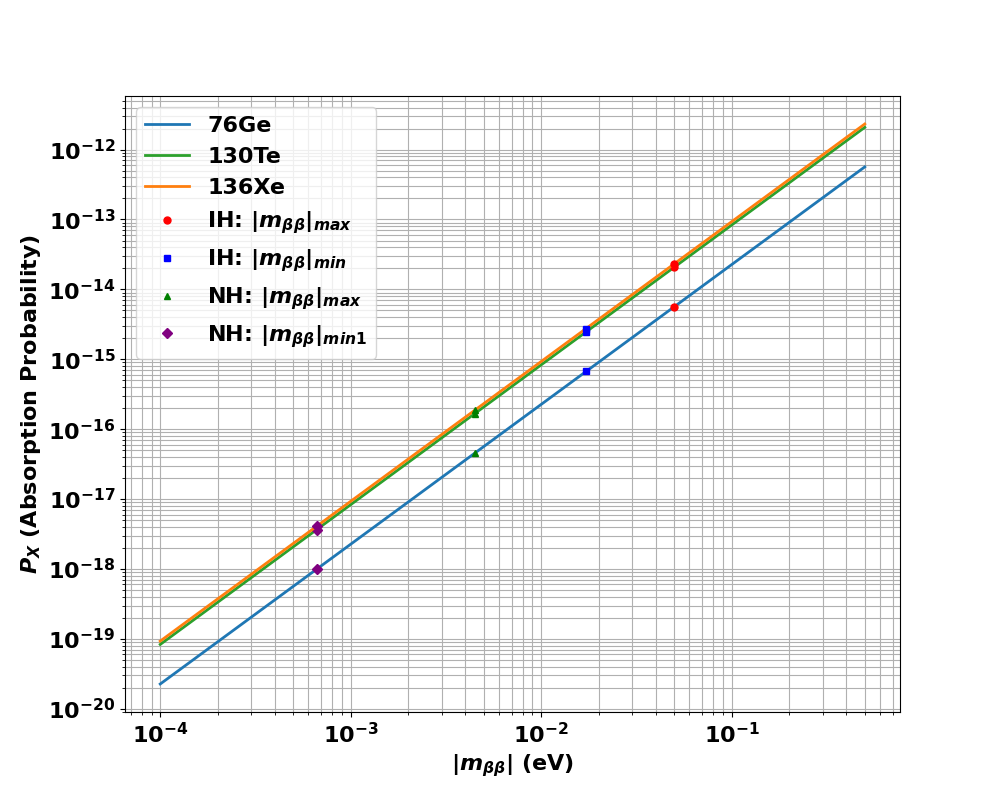}
    \caption{Absorption probability as a function of \(|m_{\beta\beta}|\) for three selected isotopes. The plot illustrates how the absorption probability of the virtual Majorana neutrino varies with the effective Majorana neutrino mass, highlighting the sensitivity of different isotopes to \(|m_{\beta\beta}|\).}
    \label{fig:DBDProbability2025}
\end{figure}

Using the values for \(\Delta E\) and \(E_{\nu_M}\) provided in Section~\ref{sec:I} for the three selected isotopes, Figure~\ref{fig:DBDProbability2025} illustrates the absorption probability as a function of $|m_{\beta\beta}|$. 

%Assuming an effective Majorana neutrino mass of \(|m_{\beta\beta}| = 0.01\,\text{eV}/c^2\), the estimated absorption (helicity-flip) probabilities for the virtual Majorana neutrino are:

%\begin{itemize}
    %\item For \(^{76}\text{Ge}\): \(P_X \approx 2.3 \times 10^{-16}\),
    %\item For \(^{130}\text{Te}\): \(P_X \approx 8.4 \times 10^{-16}\),
    %\item For \(^{136}\text{Xe}\): \(P_X \approx 9.3 \times 10^{-16}\).
%\end{itemize}

Note that the absorption probability can also be used to estimate the decay half-life through the following relation:

\begin{equation}
(T_{1/2}^{0\nu})^{-1} = G_{0\nu} \cdot |M_{0\nu}|^2 \cdot P_{X}.
\label{eq:half-life1}
\end{equation}

\begin{figure}[h]
    \centering
    \includegraphics[width=1.0\linewidth]{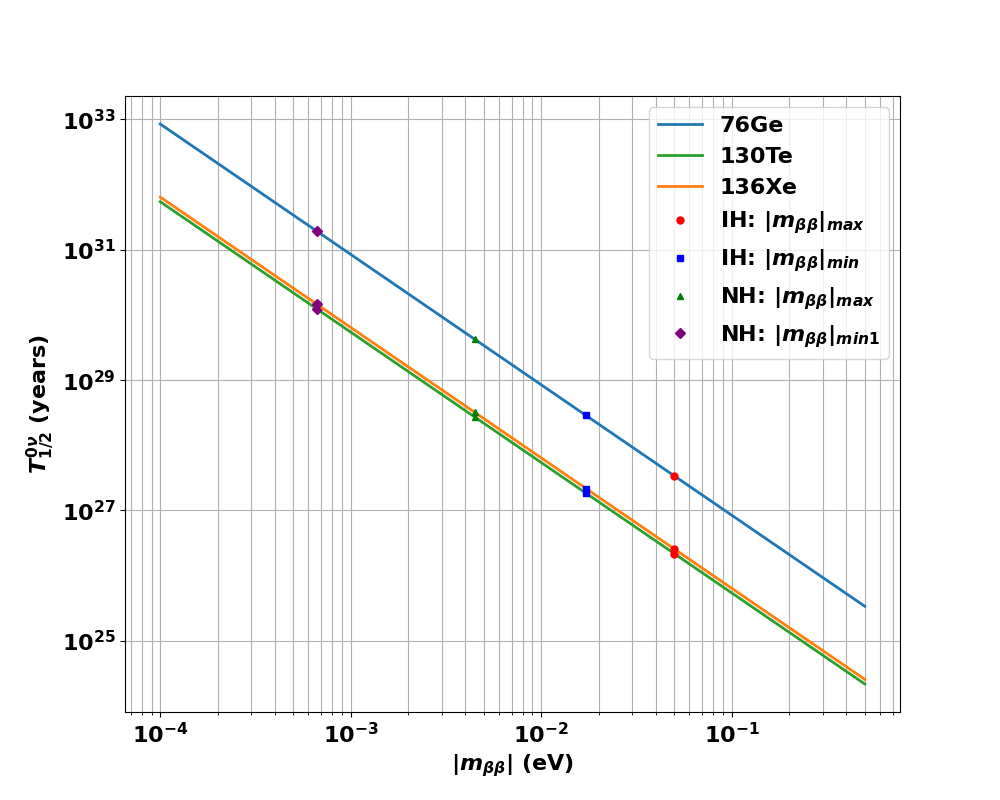}
    \caption{Decay half-life as a function of \(|m_{\beta\beta}|\). The plot illustrates the relationship between the effective Majorana neutrino mass and the predicted half-life of 0$\nu\beta\beta$ decay, highlighting the sensitivity of different isotopes to \(|m_{\beta\beta}|\).}
    \label{fig:DBDHalfLife2025}
\end{figure}

By incorporating the values of \(P_{X}\) for \(^{76}\)Ge, \(^{130}\)Te, and \(^{136}\)Xe from the calculations above, along with the corresponding values of \(G_{0\nu}\) and \(M_{0\nu}\) listed in Table~\ref{tab:nuclear_params}, we determine the respective decay half-lives. Figure~\ref{fig:DBDHalfLife2025} illustrates the relationship between the decay half-life and \(|m_{\beta\beta}|\), highlighting the sensitivity of different isotopes to variations in the effective Majorana neutrino mass.

 %By utilizing the values of $P_{X}$ for $^{76}$Ge, $^{130}$Te, and $^{136}$Xe from the calculations above, along with the corresponding values of $G_{0\nu}$ and $M_{0\nu}$ listed in Table I, we obtain the following decay half-lives:

%\begin{itemize}
    %\item $^{76}\text{Ge}$: $8.3 \times 10^{28}$ years
    %\item $^{130}\text{Te}$: $5.4 \times 10^{27}$ years
    %\item $^{136}\text{Xe}$: $6.4 \times 10^{27}$ years
%\end{itemize}

The values presented in Figure~\ref{fig:DBDHalfLife2025} closely align with the calculations shown in Figure~\ref{fig:halflife_vs_mL}, offering an alternative method for evaluating the half-life of \(0\nu\beta\beta\) decay.  This exceptionally long half-life highlights the strong dependence of left-handed neutrino absorption on the small mass of the virtual Majorana neutrino, further emphasizing why \(0\nu\beta\beta\) decay is an exceedingly rare process.

To calculate this small mass, \( |m_{\beta\beta}| \), it is essential to understand the three mass eigenstates and the total neutrino mass sum. Using the best-fit values from neutrino oscillation experiments, one can express the sum of neutrino mass eigenstates as  
    $\Sigma = m_1 + m_2 + m_3$, 
as a function of the lightest neutrino mass. This analysis reveals two distinct mass bands: one centered at approximately \(0.06\,\mathrm{eV}/c^2\) for NH and another at \(0.102\,\mathrm{eV}/c^2\) for IH. These bands enable the numerical determination of individual mass eigenstates in both scenarios, thereby establishing a possible absolute neutrino mass scale—a fundamental parameter for understanding neutrino properties and their role in the evolution of the universe.

Neutrino oscillation data, combined with cosmological constraints (such as the Planck limit of \(\Sigma < 0.072\,\mathrm{eV/c^2}\) at 95\% confidence~\cite{DESI2024}), generally favor the NH scenario. This conclusion is further supported by recent measurements from NOvA~\cite{nova} and T2K~\cite{t2k}, which have observed asymmetries in neutrino and antineutrino oscillation probabilities consistent with NH. However, while the reliability of the cosmological constraints has been questioned~\cite{sch} and individual experimental results from NOvA and T2K lack strong statistical significance, a combined analysis from these experiments has also indicated a preference for IH~\cite{denton, kevin}. Moreover, combining electron neutrino (\(\nu_e\)) disappearance data from reactor experiments with muon neutrino (\(\nu_\mu\)) disappearance data from accelerator-based experiments provides additional evidence that supports the NH scenario~\cite{hiro, step}. Overall, the determination of the neutrino mass hierarchy remains complex, and further experiments are required for a conclusive resolution.

The calculated effective Majorana mass, \(|m_{\beta\beta}|\), and the estimated \(0\nu\beta\beta\) half-lives for key isotopes such as \(^{76}\mathrm{Ge}\), \(^{130}\mathrm{Te}\), and \(^{136}\mathrm{Xe}\) provide additional means to distinguish between NH and IH. Although current experimental limits—such as GERDA’s \(1.8 \times 10^{26}\) yr for \(^{76}\mathrm{Ge}\), KamLAND-Zen’s \(2.3 \times 10^{26}\) yr for \(^{136}\mathrm{Xe}\), and CUORE’s \(2.2 \times 10^{25}\) yr for \(^{130}\mathrm{Te}\)—have not yet observed \(0\nu\beta\beta\) decay, next-generation experiments (e.g., LEGEND~\cite{LEGEND2020}, nEXO~\cite{nEXO2018}, and CUPID~\cite{CUPID2019}) aim to probe half-lives beyond \(10^{28}\) yr, reaching sensitivities that could fully test the IH scenario. To access the NH region, however, a 100-ton scale experiment is required, as described in Mei et al.~\cite{Mei2024}.

A nonzero \(|m_{\beta\beta}|\) in the NH scenario would have profound implications, suggesting that the effective Majorana neutrino mass lies within the detection range of upcoming large-scale experiments. Observation of \(0\nu\beta\beta\) decay would not only confirm that neutrinos are Majorana particles but also provide direct evidence of lepton-number violation by two units, with important consequences for theories of leptogenesis and the origin of the matter-antimatter asymmetry in the universe.

\section{Conclusion}
\label{sec:VI}
In summary, the virtual Majorana neutrino plays a pivotal role in \(0\nu\beta\beta\) decay by mediating both the transformation of an antineutrino into a neutrino and the helicity flip necessary for lepton number violation. This process is enabled by the unique properties of Majorana neutrinos, which inherently mix right-handed and left-handed components in their wavefunctions. The helicity flip is achieved through the overlap of these components, with the predominantly right-handed overlapping a highly suppressed left-handed component. This suppression, governed by the factor \(\left(\frac{|m_{\beta\beta}c^2|}{E}\right)\), directly links the decay amplitude to the neutrino mass. Moreover, the quantum entangled state of the virtual Majorana neutrino allows for wavefunction overlaps that transcend conventional energy-momentum constraints, providing a coherent mechanism for both the identity transformation and helicity flip. Thus, the virtual Majorana neutrino not only mediates \(0\nu\beta\beta\) decay but also serves as a critical link between observed lepton number violation and the fundamental Majorana nature of neutrinos, offering profound insights into the origin of neutrino masses and physics beyond the Standard Model.

The determination of the minimum neutrino mass, derived from the best-fit parameters of neutrino oscillation experiments, offers a robust pathway to understanding fundamental neutrino properties. By analyzing the sum of the neutrino masses,
$\Sigma = m_1 + m_2 + m_3$,
as a function of the lightest neutrino mass, we identify two narrow bands: one centered at approximately \(0.06\,\mathrm{eV}/c^2\) for the NH and another at \(0.102\,\mathrm{eV}/c^2\) for the IH. For NH, this corresponds to \(m_1 \approx 0.001186\,\mathrm{eV/c^2}\), while for IH, \(m_3 \approx 0.002646\,\mathrm{eV/c^2}\). These calculations confirm that the lightest neutrino mass is nonzero, which in turn implies that the effective Majorana mass \(|m_{\beta\beta}|\) must also be nonzero as depicted in Figure~\ref{fig:Majorana_mass_vs_mL}.

A nonzero \(|m_{\beta\beta}|\) is a necessary condition for \(0\nu\beta\beta\) decay if neutrinos are indeed Majorana particles. Predicted decay half-lives \(T_{1/2}^{0\nu}\) range from \(10^{27}\) to \(10^{28}\) years for IH and from \(10^{29}\) to \(10^{31}\) years for NH, highlighting the experimental challenge of detecting this rare process. Current experimental limits from GERDA, KamLAND-Zen, and CUORE are beginning to probe the IH region, but the sensitivity required to explore the NH region remains several orders of magnitude beyond current capabilities.

Planned ton-scale experiments, such as LEGEND, nEXO, and CUPID, are poised to fully explore the IH parameter space, while future 100-ton-scale detectors may be necessary to test the NH scenario. These ambitious efforts promise to provide a definitive test of the Majorana nature of neutrinos and to reveal lepton number violation, with profound implications for our understanding of the universe's origins and the mechanisms underlying leptogenesis.

In conclusion, while current experimental data slightly favor the NH scenario, a definitive resolution of the neutrino mass hierarchy and the Majorana nature of neutrinos will require significant advancements in experimental sensitivity and precision. The synergy between \(0\nu\beta\beta\) decay experiments, cosmological observations, and neutrino oscillation measurements is essential for unraveling these fundamental mysteries, which have far-reaching implications for both particle physics and cosmology.

\section{Acknowledgement} 
  This work was supported in part by NSF OISE 1743790, NSF PHYS 2310027, DOE DE-SC0024519, DE-SC0004768,  and a research center supported by the State of South Dakota.

\end{document}